\documentclass[pra,twocolumn,showpacs]{revtex4-1}
\usepackage{amsmath}
\usepackage{amssymb}
\usepackage{graphicx}
\usepackage{color}

\def\braket#1{\left\langle#1\right\rangle}

\newcommand{\gper}{\gamma_\perp}

\begin{document}

\title{General linewidth formula for steady-state multimode lasing in arbitrary cavities}

\author{Y.~D.~Chong}
\email{yidong@ntu.edu.sg}

\affiliation{Division of Physics and Applied Physics, School of Physical and Mathematical Sciences, Nanyang Technological University, Singapore 637371, Singapore}

\author{A.~Douglas Stone}

\affiliation{Department of Applied Physics, Yale University, New
  Haven, Connecticut 06520}

\begin{abstract}
A formula for the laser linewidth of arbitrary cavities in the
multimode non-linear regime is derived from a scattering analysis of
the solutions to semiclassical laser theory.  The theory generalizes
previous treatments of the effects of gain and openness described by
the Petermann factor.  The linewidth is expressed using quantities
based on the non-linear scattering matrix, which can be computed from
steady-state ab initio laser theory; unlike previous treatments, no
passive cavity or phenomenological parameters are involved.  We find
that low cavity quality factor, combined with significant dielectric
dispersion, can cause substantial deviations from the
Schawlow-Townes-Petermann theory.
\end{abstract}


\maketitle

The intrinsic linewidth of a laser arises from quantum fluctuations
and would be zero in the absence of spontaneous emission.  It is the
most important property of lasers which arises from the quantization
of the electromagnetic field.  Its value depends on the properties of
the specific laser cavity and gain medium, and was first calculated in
the seminal work of Schawlow and Townes (ST), who found the famous
linewidth formula \cite{ST}
\begin{equation}
  \delta \omega_{\textrm{ST}} = \frac{\hbar\omega_0\gamma_c^2}{2P},
  \label{omegast}
\end{equation}
where $\omega_0$ is the frequency of the laser mode, $\gamma_c$ is the
linewidth of the relevant passive cavity resonance, and $P$ is the
modal output power. Note that in this formula the properties of the
gain medium are absent.

Improved theoretical analyses over the next several decades found
three multiplicative corrections to the ST formula, all of which
tended to increase the linewidth, in some cases by large factors
\cite{Henry82,HW2}.  One correction factor arises from incomplete
inversion of the gain medium, and a second one from indirect phase
fluctuations due to the instantaneous intensity change caused by
spontaneous emission (the Henry $\alpha$ factor) \cite{Henry82}.  The
third correction and the main focus of this Letter is the Petermann factor, $K$.
First discovered in the context of transverse gain-guided
semiconductor lasers \cite{Petermann} and subsequently generalized
\cite{HW2,Haus,Henry86,Siegman,HamelWoerdman,BMS,Siegman03}, this
factor arises from the non-Hermitian nature of the laser
wave equation, due to the presence of the gain medium as well as the
openness of the laser cavity (i.e. spatially non-uniform outcoupling
loss).  It always leads to an enhancement of the linewidth, even with
uniform gain and no gain-guiding.  Typically, it is calculated from
the non-orthogonal passive cavity resonances as
\begin{equation}
K = \left|\frac{\int dr \; |\varphi(r)|^2}{\int dr \;
  \varphi(r)^2}\right|^2,
  \label{petermann}
\end{equation}
where the integrals are taken over the cavity \cite{fn1}.  In effect,
the Petermann factor changes the ST linewidth by the replacement
$\gamma_c^2 \to K \gamma_c^2$.  This is a significant correction for
lasers with large outcoupling, in the range 1.1--1.6 for the
conventional semiconductor lasers studied in Ref.~\cite{HW2}.  We
shall refer to the standard theory, inclusive of the Petermann factor,
as the Schawlow-Townes-Petermann (STP) theory.

The extensive and impressive literature on the Petermann factor
\cite{Petermann,HW2,Haus,Henry86,Siegman,HamelWoerdman,BMS,Siegman03,schomerus2}
has, with one major exception \cite{BMS}, only treated single-mode
lasing near threshold, neglecting the effects of spatial hole-burning.
And apart from a recent paper by Schomerus \cite{schomerus2}, the
literature has exclusively treated one-dimensional or waveguide
lasers, and thus is not directly applicable to the wide variety of
complex laser cavities developed during the past twenty years, such as
microdisk and deformed-disk, photonic crystal, and random lasers.  In
this Letter, we derive a general formula for the intrinsic laser
linewidth in arbitrary cavities, which is valid far from threshold,
with strong spatial hole-burning, and in the multimode regime.  The
formula relates the linewidth to a non-linear self-consistent
scattering matrix ($S$-matrix), and is based on the recently-developed
Steady-state Ab initio Laser Theory (SALT)
\cite{salt1,salt2,salt3,spasalt}.

SALT is a method for solving the steady-state properties of arbitrary
lasing structures, without directly integrating the semiclassical
laser equations.  ``Semiclassical'' here refers to the fact that the
field is treated via the classical Maxwell equations, whereas the
properties of the gain medium are obtained from a quantum-mechanical
calculation of a multi-level atom.  SALT treats the openness of the
cavity exactly, and the non-linear modal interactions and gain
saturation are included to infinite order.  Its results agree well
with numerical integration of the laser equations, but it is
computationally much more efficient \cite{opex1,opex2}.  It has been
applied to complex laser structures such as random \cite{salt3} and
photonic crystal lasers \cite{chua}.  We shall show that the quantum
input-output theory of Refs.~\cite{Caves,Loudon} can be used to
calculate quantum fluctuation properties \textit{ab initio}, in terms
of quantities obtainable from SALT.  SALT associates each laser mode
with a scattering pole---an eigenstate of a classical nonlinear
$S$-matrix with infinite eigenvalue---at a real frequency.  We derive
a formula for the linewidths of a multimode laser in terms of the
residues of these poles and a certain norm of the lasing eigenstate.
For a low-Q cavity, the generalized linewidth formula typically finds
substantial deviations from the STP prediction: typically the
linewidth is significantly {\it less} than the standard theory
predicts, and in the random laser example shown below, the laser has
an anomalous power-dependence near threshold.

The multimode SALT equations are \cite{spasalt}:
\begin{align}
  \begin{aligned}
  \left[\nabla^2 + \left(\epsilon_c(\vec{r}) + \frac{\gper
      D(\vec{r})}{\omega_\mu - \omega_a + i\gamma_\perp}
    \right)\omega_\mu^2\right] \Psi_\mu(\vec{r}) = 0, \\
  D(\vec{r}) = D_0(\vec{r}) \, \left[1 +\sum_{\nu=1}^n \Gamma_\nu
    |\Psi_\nu(\vec{r})|^2\right]^{-1}, \label{TSG}
  \end{aligned}
\end{align}
where $\Psi_\mu$ is the $\mu^{\textrm{th}}$ steady-state lasing mode,
$\omega_\mu$ is its frequency, $\epsilon_c$ is the passive cavity
dielectric function, $\gper$ is the gain medium linewidth, $\omega_a$
is the atomic transition frequency, $D_0 (\vec{r})$ is the (possibly
spatially-varying) pump, and $\Gamma_\nu \equiv \gper^2/(\gper^2 +
(\omega_\nu-\omega_a)^2)$ is the gain curve.  The effective pump
$D(\vec{r})$ contains an infinite-order nonlinear ``hole-burning''
term, which gives rise to mode competition and gain saturation in a
quantitatively precise manner. These coupled, time-independent,
non-linear equations are solved with the boundary condition of purely
outgoing waves with frequency $\omega_\mu$ at infinity; the solution
algorithm is discussed in Refs.~\cite{salt3,spasalt,opex2}.

From the solution to (\ref{TSG}), we can compute a self-consistent
$S$-matrix for any complex frequency $\omega$, not just the discrete
lasing frequencies $\omega_\mu$.  By definition,
this $S$-matrix has one or more poles on the real-$\omega$ axis, at
$\omega = \omega_\mu$.  It can be used to study the effects of vacuum
fluctuations and spontaneous emission \cite{fn2}.  Suppose the cavity
has scattering channels indexed by $j = 1,2,\dots,N$ (e.g.~waveguide
modes or spherical waves, depending on the scattering geometry).  The
input and output photon operators, denoted by $a_1,\dots,a_N$ and
$b_1,\dots,b_N$ respectively, obey an ``input-output'' relation
\cite{beenakker0}:
\begin{equation}
  b_i(\Omega) = \sum_j S_{ij}(\Omega) \, a_j(\Omega) + \sum_\rho
  V_{i\rho}(\Omega)\, d_\rho^\dagger(-\Omega).
  \label{input output}
\end{equation}
Here the frequency $\Omega$ is measured from the lasing
frequency of interest, which we denote by $\omega_0$, $\Omega \equiv \omega - \omega_0$.  The
$d_\rho$'s are ladder operators for the external reservoirs
corresponding to the gain medium, with the index $\rho$ denoting
appropriate degrees of freedom in the cavity/reservoir \cite{fn3}.  

In order for $a$, $b$, and $d$ to obey canonical commutation
relations, e.g.~$[a_i(\Omega),a_j(\Omega')] =
\delta_{ij}\,\delta(\Omega-\Omega')$, the $S$-matrix must be related
to the reservoir coupling coefficients by the fluctuation-dissipation
relation \cite{beenakker0}
\begin{equation}
  SS^\dagger - VV^\dagger = \mathbf{1},
  \label{fluct-diss}
\end{equation}
where $\mathbf{1}$ is the $N\times N$ identity matrix.  Next, we
define
\begin{equation}
  a_j(t) = \frac{1}{\sqrt{2\pi}} \int d\Omega\; a_j(\Omega) \;
  e^{-i\Omega t}, \label{at_def}
\end{equation}
and similarly for $b_j(t)$ and $d_\rho(t)$, describing quantum
amplitudes for the field envelopes.  Inserting into (\ref{input
  output}) gives
\begin{align}
  \begin{aligned}
    b_i(t) = &\int dt'\; \left[\sum_k \int \frac{d\Omega}{2\pi}\;
      S_{ik}(\Omega) \, e^{-i\Omega(t-t')} \right] a_k(t') \\ + & \int
    dt'\; \left[\sum_\rho \int \frac{d\Omega}{2\pi}\;
      V_{i\rho}(\Omega) \, e^{-i\Omega(t-t')} \right]
    d_\rho^\dagger(t').
    \label{at2}
  \end{aligned}
\end{align}
The first term describes scattering of input photons, and the second
describes emission from the gain medium.

$S$ is strongly constrained by its symmetries.  Firstly, optical
reciprocity \cite{reciprocity} implies that $S$ can be written as a
symmetric matrix, so it has the eigenvalue decomposition
\begin{equation}
  S = \sum_n |\psi_n\rangle\,
  \frac{s_n}{\langle\psi_n^*|\psi_n\rangle}\, \langle\psi_n^*|,
  \label{S expansion}
\end{equation}
where each $|\psi_n\rangle$ denotes a right eigenvector of $S$ with
eigenvalue $s_n$, and $\langle\psi_n^*|$ denotes its unconjugated
transpose.  These eigenvectors are bi-orthogonal ($\langle\psi_m^*|
\psi_n\rangle = 0$ for $m \ne n$) and power-normalized ($\langle
\psi_n| \psi_n\rangle = 1$).

Suppose that $\epsilon_c(r)$ is real.  The $S$-matrix of the passive
cavity is unitary, and for a high-Q cavity with a resonance near
$\omega_0$, one of the eigenvalues is approximately \cite{Walls}:
\begin{equation*}
  s_0(\Omega) \approx e^{i\varphi(\Omega)} \; \frac{\Omega -
    i\gamma_c/2}{\Omega + i\gamma_c/2},
\end{equation*}
where $\varphi$ is an irrelevant phase factor and $\gamma_c$ is the
cavity lifetime.  The eigenvalue is unimodular for real $\Omega$, and,
as required by time-reversal symmetry, its poles and zeros lie at
conjugate positions in the complex $\Omega$ plane.

Adding gain pushes the zero and pole up in the complex frequency
plane.  The eigenvalue takes the form
\begin{equation}
 s_0(\Omega) \approx e^{i\varphi'(\Omega)} \; \frac{\Omega - i \Gamma_z}{\Omega + i \Gamma_p},
  \label{s0}
\end{equation}
where $\Gamma_z$ and $\Gamma_p$ are the distances of the zero and pole
from the real axis.  The lasing threshold is reached as $\Gamma_p
\rightarrow 0^{-}$; within the high-Q approximation the eigenvalue
takes the form (\ref{s0}) with $\Gamma_z \approx \gamma_c$ (the zero
moves up the same distance as the pole).  This approximation leads
directly to the ST formula (high Q will imply $K \approx 1$).  For
arbitrary Q, the $S$-matrix near $\Omega = 0$ takes the form
(\ref{s0}), with a generalized residue $\Gamma_L (\Omega)$ replacing
$\Omega - i \Gamma_z$.  We denote the eigenvector corresponding to
this diverging eigenvalue by $\Psi$.  In the $S$-matrix decomposition
(\ref{S expansion}), the term with $s_0$ dominates, so we can write
\begin{equation}
  S \approx |\Psi\rangle \, \frac{s_0}{\langle\Psi^*|\Psi\rangle}\,
  \langle\Psi^*|.
  \label{S projection}
\end{equation}
Using this together with Eq.~(\ref{fluct-diss}) gives
\begin{equation}
  VV^\dagger \approx |\Psi\rangle \frac{1}{|\Psi^T\Psi|^2} \;
  \frac{|\Gamma_L|^2}{\Omega^2 + \Gamma_p^2} \langle \Psi|.
\end{equation}
This equation is satisfied by the ansatz
\begin{equation}
  V_{i\rho} = \frac{1}{\Psi_L^T\Psi_L} \; \frac{\Gamma_L}{\Omega + i
    \Gamma_p} \; \Psi_L^i \, u_\rho,
  \label{basic V ansatz}
\end{equation}
where $u$ is some vector satisfying $\sum_\rho u^\dagger_\rho u_\rho =
1$, and $\Psi_L^i$, is the $i^{\textrm{th}}$ component of the
$S$-matrix eigenvector for the lasing mode.  Note that this relation
applies not just to the first lasing mode at threshold, but also for
above-threshold steady-state lasing modes, using the self-consistent,
non-linear $S$-matrix obtained from SALT.

Inserting (\ref{basic V ansatz}) into (\ref{at2}) and performing the
resulting contour integrals gives
\begin{eqnarray}
  b_i(t) &=& - \frac{\Gamma_L \Psi_L^i}{\Psi_L^T\Psi_L} \int^t dt'
  e^{-\Gamma_p (t-t')} F(t') \label{bit}
  \\ F(t) &\equiv& \sum_j \Psi_{Lj} \,a_j(t) + i \sum_\rho u_\rho\,
  d_\rho^\dagger(t). \label{Ft}
\end{eqnarray}
Thus each output photon is a superposition of incoming photons and
reservoir excitations from all \textit{earlier} times.

Above threshold, the gain medium undergoes stimulated emission, and
the laser field acquires a mean value, $B_i$, so that Eq.~(\ref{bit} becomes:
\begin{equation}
  b_i(t) = B_i - \frac{\Gamma_L \Psi_L^i}{\Psi_L^T\Psi_L} \int^t dt' e^{-\Gamma_p
    (t-t')} F(t'),
  \label{bit2}
\end{equation}
where $B_i$, the steady-state classical outgoing
field amplitude in channel $i$, is related to $\Psi_L^i$ by
\begin{eqnarray}
  |B_i|^2 = \frac{P}{\hbar \omega_0} \, \left|\Psi_L^i\right|^2
  \label{total power}
\end{eqnarray}
where $P$ is the total output power of the mode.

Due to the fluctuation operator $F(t)$, the phase of the output field
has a quantum uncertainty; the rate at which this uncertainty
increases with time gives the laser coherence time scale.  The
fluctuation-induced phase changes are fed back into the classical
value of $B_i$, causing a random drift in the phase of the laser
field.  We ignore this feedback, instead taking a fixed value for
$B_i$ for all $t$.  This is justifiable because the integrand in
(\ref{bit2}) vanishes exponentially for $t' \lesssim - T$, where $T =
1/\Gamma_p$ will turn out to be the coherence time.  The calculations
below apply to times much shorter than $T$.

\begin{figure}
\centering
\includegraphics[width=0.44\textwidth]{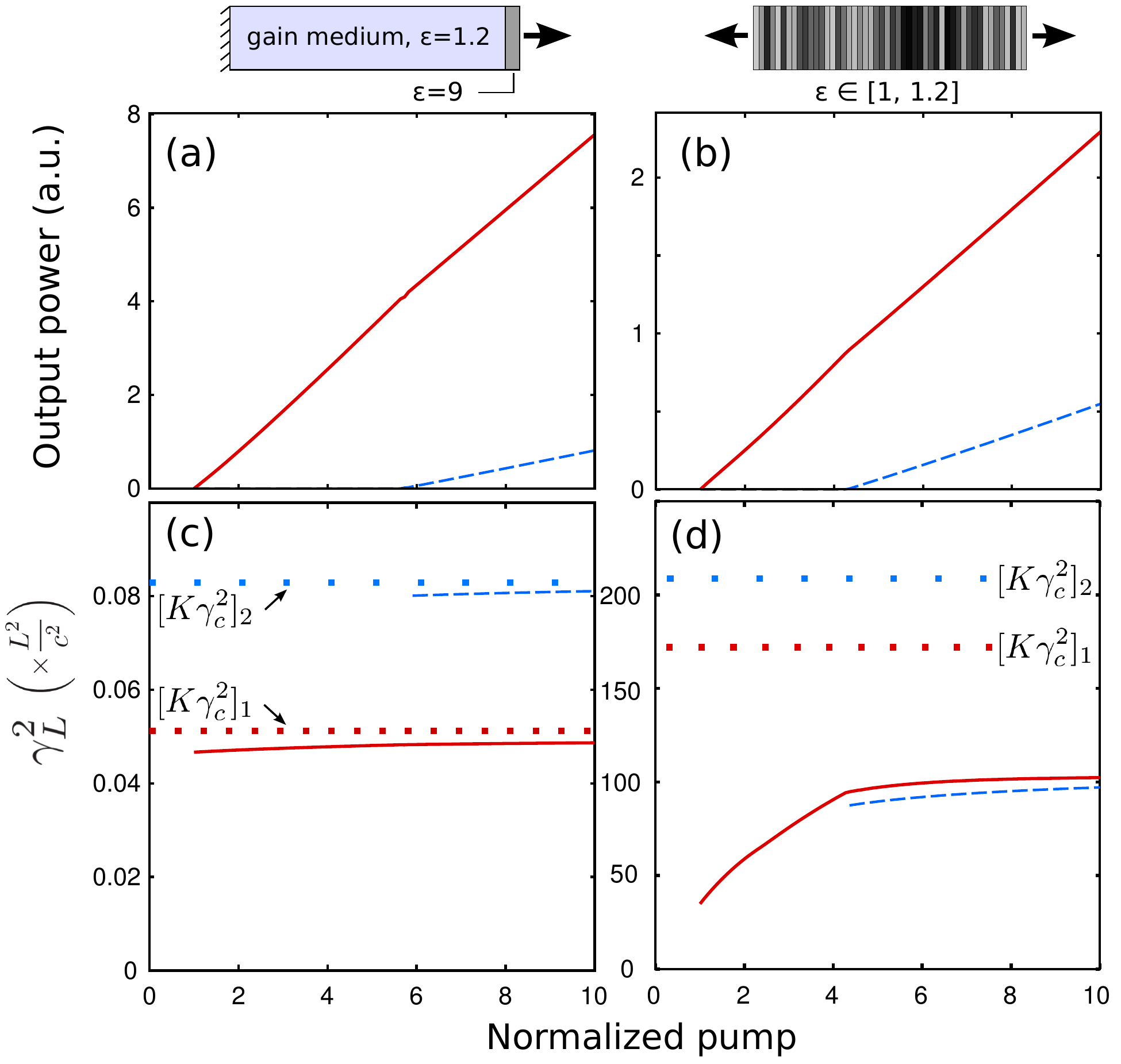}
\caption{(color online) Output power and cavity decay rates
  $\gamma_L^2$ for two uniformly pumped one-dimensional microcavity
  lasers.  A slab of gain material with background $\epsilon = 1.2$,
  bounded on the left by a perfect mirror and on the right by an
  $\epsilon = 9$ slab (5\% of the total length) acting as a partially
  transmitting mirror (left schematic).  A random laser consisting of
  50 slabs of gain material, each with background $\epsilon$ uniformly
  distributed in $[1, 1.2]$ (right schematic).  Both systems exhibit
  two-mode lasing at the high end of the pump range.  Plots (a),(b)
  show modal output powers \textit{vs.}~the normalized pump
  \cite{spasalt}.  Plots (c),(d) show the square of the generalized
  cavity decay rate $\gamma_L^2 \equiv
  |\Gamma_L|^2/|\Psi_L^T\Psi_L|^2$ which determines the linewidths
  according to Eq.~(\ref{linewidth result}).  Solid and dashed curves
  denote modes 1 and 2 respectively.  The horizontal dotted lines show
  the conventional result, $K \gamma_c^2$, computed from the passive
  cavity quasimodes, which fails for the random laser. }
\label{fig_residues}
\end{figure}

We choose the arbitrary global phase of $B_i$ to be
real and positive for the specific channel $i$, and study the quantum
fluctuations of the phase via the Hermitian quadrature
operator \cite{Clerk}
\begin{equation}
  \theta_i = \frac{i(b_i^\dagger - b_i)}{2B_i},
\end{equation}
which for small phase angles corresponds to the phase of the laser
output in channel $i$.  Using (\ref{Ft}) and (\ref{bit2}), we compute
the quantity $\braket{\theta_i(t_1)\theta_i(t_2)}$, taking $\braket{a}
= \braket{d} = \langle a^\dagger_i(t_1) \, a_j(t_2)\rangle = 0$ and
taking the white noise correlator
\begin{eqnarray}
  \braket{d_\rho(t_1) \, d_\nu^\dagger(t_2)} = f_\rho \,
  \delta_{\rho\nu} \, \delta(t_1-t_2),
\end{eqnarray}
where $f_\rho = [P_2 / (P_2 - P_1)]_\rho$ describes the local
population inversion \cite{beenakker0}.  The zero-point contributions
to $\braket{\theta_i(t_1)\theta_i(t_2)}$ from the photon input and the
gain medium cancel exactly, leaving
\begin{equation}
  \braket{\theta_i(t_1)\theta_i(t_2)} =
  \frac{|\Gamma_L|^2}{|\Psi_L^T \Psi_L|^2}\,
  \frac{\hbar\omega_0}{4\Gamma_pP} \; e^{-\Gamma_p|t_1-t_2|}
  \; \bar{f},
  \label{theta theta}
\end{equation}
where $P$ is the modal output power from Eq.~(\ref{total power}), and
\begin{equation}
  \bar{f} \equiv \sum_\rho f_\rho |u_\rho|^2
\end{equation}
is the inversion factor correction mentioned at the beginning of this
Letter.

\begin{figure}
\centering
\includegraphics[width=0.45\textwidth]{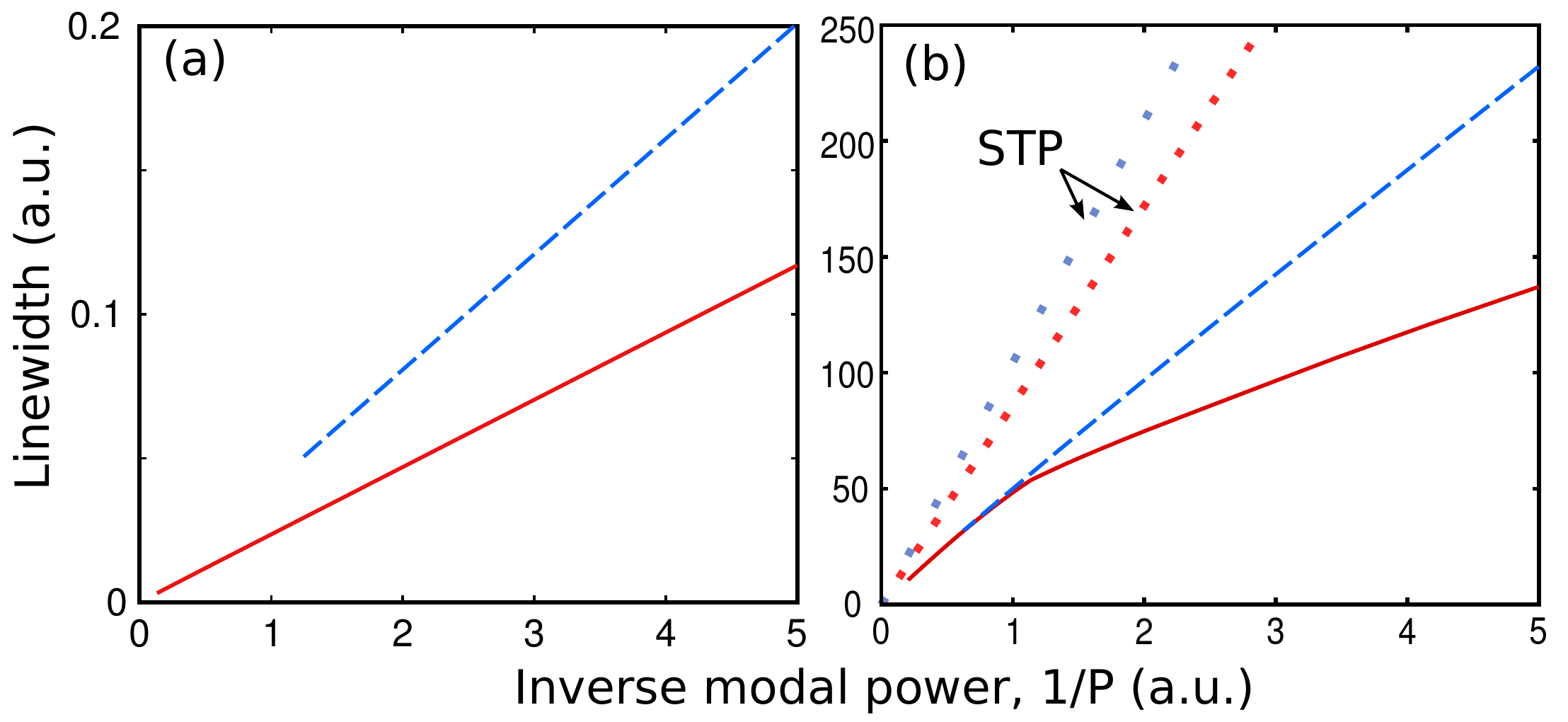}
\caption{(color online) Laser linewidth $\textit{vs.}$ inverse modal
  output power $1/P$, for the two lasers studied in
  Fig.~\ref{fig_residues}.  The linewidths are computed using
  Eq.~(\ref{linewidth result}), assuming the inversion factor
  $\bar{f}=1$.  (a) The high-Q cavity laser linewidths show the
  standard $1/P$ dependence for both modes. (b) The linewidth of the
  first mode of the random laser deviates strongly from the $1/P$
  Schawlow-Townes-Petermann dependence at lower pump values.  At large
  pump values the linewidths of both mode 1 (solid curve) and mode 2
  (dashed curve) vary as $1/P$, but with values roughly half that of
  the standard STP prediction (dotted curve).  }
\label{fig_linewidths}
\end{figure}

The phase uncertainty accumulated over time $\Delta t$ is
$\braket{[\theta_i(t+\Delta t)-\theta_i(t)]^2} = \Delta \omega \Delta
t + O(\Delta t^2)$, where
\begin{equation}
  \Delta \omega =  \frac{|\Gamma_L|^2}{|\Psi_L^T \Psi_L|^2}\,
  \frac{\hbar\omega_0}{2P} \, \bar{f} \equiv  
  \frac{\hbar\omega_0 \gamma_L^2}{2P} \, \bar{f}
   \label{linewidth result}
\end{equation}
This is our central result: a general linewidth formula in which
$|\Gamma_L|^2/|\Psi_L^T \Psi_L|^2 \equiv \gamma_L^2$ replaces the
quantity $K \gamma_c^2$ in the conventional Schawlow-Townes-Petermann
linewidth formula.  We can think of $\gamma_L$ as a generalized cavity
decay rate, corrected for the presence of gain, openness,
hole-burning, and saturation.  It is calculable \textit{ab initio},
with no phenomenological parameters, from the non-linear classical
$S$-matrix of SALT. The lasing eigenvector is found by diagonalizing
the $S$-matrix at each lasing pole, and the residue $\Gamma_L$ is
found by numerically integrating the relevant eigenvalue of the
$S$-matrix around the pole.  Eq.~(\ref{linewidth result}) only
includes the contribution to the laser linewidth from direct phase
fluctuations; the indirect phase fluctuations \cite{Henry82} have been
omitted for simplicity.

The relation of the Petermann factor to the residue of the lasing pole
for a waveguide laser was emphasized early on by Henry \cite{Henry86},
and developed for more general cavities in an $S$-matrix formulation
in Refs.~\cite{schomerus,schomerus2}, but in all previous cases for a
single lasing mode at threshold, i.e.~without non-linear effects.
Goldberg \textit{et al.}~\cite{BMS} gave an excellent and detailed
analysis of the linewidth for multimode lasing, including non-linear
effects, but using an approach applicable only to one-dimensional
cavities with spatially uniform dielectric functions.  To our
knowledge, our Eq.~(\ref{linewidth result}), combined with SALT, is
unique in providing a quantitative method for calculating the
intrinsic laser linewidth in arbitrary cavities and pump profiles in
the multimode, non-linear regime.  Assuming steady-state multimode
lasing exists, the present theory makes no significant further
approximations, and hence it can be used to evaluate the validity of
the STP linewidth formula \cite{HW2}.

We can connect Eq.~(\ref{linewidth result}) to previous results
involving quasi-modes, such as Refs.~\cite{Petermann,Haus,Siegman}, by
examining the $S$-matrix of a \textit{passive} cavity.  A quasimode
$\varphi(r)$ is a purely-outgoing solution to the wave equation for a
passive cavity with dielectric function $\epsilon(r)$, at complex
frequency $\omega_p$, where $\textrm{Im}(\omega_p) \equiv
-\gamma_c/2$.  Let $\Psi$ be the $S$-matrix eigenvector for this pole,
normalized by $\Psi^\dagger\Psi = 1$, and let $\Gamma$ be the residue
of the eigenvalue.  It can be shown that
\begin{eqnarray}
  \,\textrm{Im} \left[\omega_p^2 \int dr \; \epsilon(r) \,
    |\varphi(r)|^2\right] &=& - \textrm{Re}[\omega_p], \label{residue
    relation 1} \\ \int \epsilon(r) \, \varphi^2(r) &=& \left[
    \frac{i}{\Gamma} -\frac{i}{2\omega_p} \right] \Psi^T\Psi.
  \label{residue relation 2}
\end{eqnarray}
Here the spatial integrals are taken over the cavity.  For real
$\epsilon(r)$, and in the limit $|k_p| \gg \Gamma \sim \gamma_c$,
(\ref{residue relation 1})-(\ref{residue relation 2}) give
\begin{equation}
  \frac{|\Gamma|^2}{|\Psi^T\Psi|^2} \;\approx\; \left|\frac{\int dr \;
    \epsilon(r) \; |\varphi(r)|^2}{\int dr \; \epsilon(r) \;
    \varphi(r)^2}\right|^2 \gamma_c^2 = K \gamma_c^2.
  \label{pole wavefunction relation}
\end{equation}
Thus, in this slightly generalized version, $K \gamma_c^2$ is
approximately equal to our $\gamma_L^2$, when evaluated for the
passive cavity.  Note that both (\ref{pole wavefunction relation}) and
its active-cavity generalization in Ref.~\cite{schomerus2} do not
include the effects of dielectric dispersion, which can have a
significant effect on $\gamma_L^2$.

Fig.~\ref{fig_residues} compares $\gamma_L^2$ to $K \gamma_c^2$ for
two one-dimensional microcavity multimode lasers: a high-Q, uniform
cavity for which the two quantities agree rather well, and a low-Q
random laser, for which major deviations are found.  For the random
laser, at pump strengths up to four times threshold, $\gamma_L$ for
the first lasing mode depends strongly on $P$, causing the overall
power dependence to depart substantially from the standard $1/P$
dependence (Fig.~\ref{fig_linewidths}). For higher pump strengths,
$\gamma_L$ is approximately constant, but the conventional linewidth
prefactor $K \gamma_c^2$ {\it overestimates} it by almost a factor of
two.  In the standard theory, the STP linewidth is expected to be a
lower bound set by field quantization, but insofar as the usual STP
formula relies on passive cavity quantities it is not a reliable
bound.  Analysis of our results indicates that this deviation from the
STP theory arises from low cavity Q and from the frequency dispersion
of the dielectric constant of the gain medium, which cause a
significant reduction of the residue of the lasing pole at threshold
compared to its value in the passive cavity.  We do not believe that
the apparent violation of the STP bound indicates any new quantum
fluctuation properties of the laser.  In future work, our generalized
linewidth formula will allow such issues to be studied systematically.

This research was partially supported by NSF grant No.~DMR-0908437,
and by NRF (Singapore) grant No. NRFF2012-02.  The authors would like
to thank H. Cao, M. Devoret, and H.~Schomerus for helpful discussions.



\begin{thebibliography}{99}
\bibitem{fn1} In Eq.~(24) we show that a more accurate formula for the
  Petermann factor of the passive cavity includes the dielectric
  function $\epsilon (r)$ in the integrands.
\bibitem{fn2} We will assume that the quantum fluctuations are small
  compared to the average classical field, so we can neglect any
  additional non-linear saturation of the gain medium due to those
  effects.
\bibitem{fn3} Note that to satisfy causality these operators enter with
frequency $-\Omega$.
\end{thebibliography}
\end{document}